\documentclass[prd,nofootinbib,preprintnumbers,floatfix]{revtex4}  
\usepackage[plainpages=false, colorlinks=true, anchorcolor=blue, linkcolor=blue, citecolor=blue, bookmarks=false]{hyperref}
\usepackage{placeins}
\usepackage{array}
\usepackage{amsmath}
\usepackage{natbib}
\usepackage{tabularray}
\newcommand{\rthis}[1]{\textcolor{black}{#1}}
\usepackage{graphicx}

\newcolumntype{P}[1]{>{\centering\arraybackslash}p{#1}}
\begin{document}
\author{Aaseesh  \surname{Rallapalli}$^{1}$}%
 \altaffiliation{ee20btech11060@iith.ac.in}
 \author{Shantanu \surname{Desai}$^{2}$ }%
 \altaffiliation{shntn05@gmail.com}
\affiliation{$^{1}$ Department of Electrical Engineering, Indian Institute of Technology, Hyderabad, Kandi, Telangana-502284  India}

\title{Bayesian inference of $W$-boson mass}
\affiliation{$^{2}$ Department of Physics, Indian Institute of Technology, Hyderabad, Kandi, Telangana-502284  India}
\date{\today}

\begin{abstract}
We  use a  Bayesian regression technique (similar to a recent analysis by  Rinaldi et al) to  obtain a central estimate for   the $W$-boson mass using four different combinations of  datasets compiled by the PDG including the 2022 CDF result. We use three different priors on the unknown intrinsic scatter and also   a non-parametric hierarchical Dirichlet Process Gaussian Mixture model to obtain a world average for  $W$-boson mass. We also evaluate the statistical significance of the  discrepancy with respect to the Standard model for each of the datasets. We find that for all the combination of datasets and the aformentioned prior choices, the discrepancy with respect to the Standard Model value for the $W$-mass is less than 3$\sigma$. We also checked that if we use a narrow prior on the intrinsic scatter, we get a discrepancy of about 3.8$\sigma$ compared to the Standard model value.
\end{abstract}
\maketitle

\section{Introduction}
The accurate determination of $W$-boson mass is of paramount importance as it provides a powerful  probe of Physics beyond Standard model~\cite{PDG}. Its mass has been determined using both $e^{+}e^{-}$ as well as $p-\bar{p}$ colliders. As of April 2022, the PDG has compiled a series of 25 measurements following the discovery of the $W$-boson in 1983. The world average of all the  $W$-boson measurements tabulated by PDG is equal to $80.377 \pm 0.012$ GeV. In April 2022, the CDF collaboration came out with a new measurement using the full Run 2 dataset with an integrated  luminosity of 8 $fb^{-1}$,  and obtained a mass estimate of  $80.4335 \pm 0.0094$ GeV~\cite{CDF22}. This result is 7$\sigma$ discrepant with respect to the standard model expectation of $80.357 \pm 0.006 $ GeV~\cite{PDG}, potentially insinuating towards new Physics~\cite{HillPT}. It also disagrees with the PDG world average at about 3.7$\sigma$. \rthis{If one combines the average of CDF measurements ($80.4335 \pm 0.0094$) GeV  along with the remaining PDG measurements ($80.377 \pm 0.012$) GeV using least squares averaging, we get a central estimate of ($80.412 \pm 0.0074$) GeV and the tension between the Standard Model value becomes  5.77$\sigma$.}
In order to reconcile the 2022 CDF result with previous measurements, the  uncertainties of all results would need to be rescaled by a factor of two in order to give a reduced $\chi^2$ of unity~\cite{PDG}.

Given the paramount importance of getting an accurate estimate of the $W$ mass ($M_W$, hereafter) as its precise measurement  potentially points to new Physics, we obtain an independent central estimate from  the $W$ measurements compiled in PDG using  Bayesian regression analysis in order to understand the systematics in the mass measurements of $W$ and to assess the statistical significance of any discrepancies with respect to the Standard Model by combining all the measurements.  The world average estimated in this way is complementary to that obtained in PDG, which has been obtained  from the  weighted mean, which  is an inherently  frequentist procedure.

This Bayesian method  which we use in this manuscript follows a  recent work by ~\citet{Rinaldi} (R22, hereafter), who applied these methods to Newton's constant ($G$) measurements in order to understand the systematic errors in the measurements which have been discrepant with respect to each other. (See also ~\cite{Bethapudi}). In addition to an estimate for $G$, R22 also accounted for an unknown systematic error.
R22 used three different priors on the systematic error,  and also used a non-parametric hierarchical Dirichlet mixture model technique to get a central estimate for $G$ together  with the associated systematic error for each prior choice. We apply the same analysis techniques  as in R22 to the $W$ mass measurements collated in PDG. \rthis{Finally, we note that similar methods for combining discrepant data using Bayesian inference have also been proposed in High energy physics literature~\cite{Agostini}.}

The outline of this manuscript is as follows. We summarize the  analysis carried out by R22 in Sect~\ref{sec:recap}.  We present the data analyzed and results in Sect.~\ref{sec:results}. We conclude in  Sect.~\ref{sec:conclusions}.

\begin{table}[!htb]
\centering
\begin{tabular}{ |P{3cm}|P{3cm}|P{3cm}|P{3cm}|P{3cm}|  }
\hline
\multicolumn{5}{|c|}{Measurements of $W$ Boson Mass compiled in PDG} \\
\hline
Measurement  & EXPT &  $M_W$ (GeV) & Uncertainties (GeV) & Used by PDG \\
\hline
AALTONEN22~\cite{CDF22} & CDF & 80.4335 & $\pm$0.0094& Used \\
AAIJ22C~\cite{Aij22} & LHCB & 80.354 & $\pm0.023$$\pm$0.022 & Used \\
AABOUD18J~\cite{aaboud18} & ATLS & 80.370 & $\pm0.007$$\pm0.017$ & Used \\
AALTONEN12E~\cite{CDF2012} & CDF & 80.387 & $\pm0.012$$\pm0.015$ & Used \\
ABAZOV12F~\cite{Abazov12} & D0 & 80.375 & $\pm0.011$$\pm0.020$ & Used \\
ABDALLAH08A~\cite{Abdallah08} & DLPH & 80.336 & $\pm0.055$$\pm0.039$ & Used \\
ABBIENDI 06~\cite{OPAL:2005rdt} & OPAL & 80.415 & $\pm0.042$$\pm0.031$ & Used \\
ACHARD06~\cite{L3:2005fft} & L3 & 80.270 & $\pm0.046$$\pm0.031$ & Used \\
SCHAEL06~\cite{ALEPH:2006cdc} & ALEP & 80.440 & $\pm0.043$$\pm0.027$ & Used \\
ABAZOV02D~\cite{D0:2002fhu} & D0 & 80.483 & $\pm0.084$ & Used \\
AFFOLDER01E~\cite{CDF:2000gwd} & CDF & 80.433 & $\pm0.079$ & Used \\
ANDREEV18A~\cite{H1:2018mkk} & H1 & 80.520 & $\pm0.070$$\pm0.092$ & Not Used \\
ABAZOV12F~\cite{D0:2012kms} & D0 & 80.367 & $\pm0.013$$\pm0.022$ & Not Used \\
ABAZOV09AB~\cite{D0:2009yxq} & D0 & 80.401 & $\pm0.021$$\pm0.038$ & Not Used \\
AALTONEN07F~\cite{CDF:2007cmy} & CDF & 80.413 & $\pm0.034$$\pm0.034$ & Not Used \\
AKTAS06~\cite{H1:2005xal} & H1 & 82.87 & $\pm1.82$$^{+0.30}_{-0.16}$ & Not Used \\
CHEKANOV02C~\cite{ZEUS:2002mim} & ZEUS & 80.3 & $\pm2.1$$\pm1.2$$\pm1.0$ & Not Used \\
BREITWEG00D~\cite{ZEUS:1999yhz} & ZEUS & 81.4 & $^{+2.7}_{-2.6}$$\pm2.0$$^{+3.3}_{-3.0}$ & Not Used \\
ALITTI92B~\cite{UA2:1991hww} & UA2 & 80.84 & $\pm0.22$$\pm0.83$ & Not Used \\
ALITTI90B~\cite{UA2:1990jit} & UA2 & 80.79 & $\pm0.31$$\pm0.84$ & Not Used \\
ABE89I~\cite{CDF:1988ocw} & CDF & 80.0 & $\pm3.3$$\pm0.84$ & Not Used \\
ALBAJAR89~\cite{UA1:1988rck} & UA1 & 82.7 & $\pm1.0$$\pm2.7$ & Not Used \\
ALBAJAR89~\cite{UA1:1988rck} & UA1 & 81.8 & $^{+6.0}_{-5.3}$$\pm2.6$ & Not Used \\
ALBAJAR89~\cite{UA1:1988rck} & UA1 & 89 & $\pm3$$\pm6$ & Not Used \\
ARINSON83~\cite{UA1:1983crd} & UA1 & 81. & $\pm5$ & Not Used \\
BANNER83B~\cite{UA2:1983tsx} & UA2 & 80. & $^{+10.}_{-6.}$ & Not Used \\

\hline
\end{tabular}
\caption{\label{tab:PDG} Summary of PDG compilations of  $W$-boson mass measurements, which are also used in this work. The first column indicates the annotation used for the measurement and is same as that used by PDG~\cite{PDG}. The last column indicates whether the measurement was used or not by PDG for calculating the world average.}
\end{table}
\section{Recap of R22}
\label{sec:recap}
The main goal of R22 was to characterize the systematic errors in measurements of $G$. For this purpose R22 used both a parametric and non-parametric Bayesian regression method. We briefly recap the analysis procedure in R22, where more details can be found. More detailed expositions on Bayesian analysis can be found elsewhere~\cite{Sharma,Krishak,Trotta,Weller}

The first goal in Bayesian parameter estimation for a parameter vector ($\theta$) given some data $D$ and model $M$ is to obtain the posterior distribution $P(\theta|D,M)$ given some model $M$. 
This can be obtained from Bayes theorem as follows:
\begin{equation}
        P(\theta|D,M) = \frac{P(D|M,\theta)P(\theta|M)}{P(D|M)}
        \label{eq:bayesthm}
    \end{equation} 
where $P(D|M,\theta)$ represents the likelihood and $P(\theta|M)$ the prior on the parameter vector $\theta$. The parameter vector  ($\theta$)  considered in R22 is given by \{$G$, $\Sigma\}$, where $\Sigma$ is the unknown systematic error (or intrinsic scatter), which  also has been kept as a free parameter. The dataset $D$ \rthis{consists of $N$ measurements of the parameter $G$, given by  $D=\{\hat{G}_1,... \hat{G}_N\}$. We model the $i^{th}$ measurement $\hat{G_i}$ as following a Gaussian distribution with mean $G_i$ and variance $\sigma_i^2$, and the $G_i$ themselves have a Gaussian prior with mean $G$ and variance $\Sigma^2$. After marginalizing over $G_i$ we get:}
\begin{equation}
    P(D|M,\theta) = \prod_i^N \int \mathcal{N}  ( \hat{G_i}| G_i,\sigma_i)  \mathcal{N} (G_i | G,\Sigma)dG_i=  \prod_i^N \mathcal{N} \left(G | \hat{G_i},\sqrt{\sigma_i^2+\Sigma^2}\right),
\label{eq:likelihood}    
\end{equation}
where  $\sigma_i$ denote the errors in the measurements of $G$;  $N$ is the total number of data points; and $\mathcal{N}$ represents the Gaussian distribution. 
\rthis{In this context, the intrinsic scatter parameter $\Sigma$ can be interpreted as an an additional systematic uncertainty assigned in each measurement, that goes beyond the quoted uncertainties. We note that if the prior on $\Sigma$ is chosen to be very broad,  Bayes' theorem adjusts it value to reflect any scatter  amongst the different measurements that goes beyond the reported measurement errors $\sigma_i$. However, because of the potential degeneracy between $G$ and $\Sigma$, one could also get broader posteriors on $G$. The exact parameter where the degeneracy gets absorbed however cannot be predicted apriori.}
The prior $P(\theta|M)$ consists of a prior on $G$ as well as a prior on the intrinsic scatter ($\Sigma$). The unknown intrinsic scatter is added in quadrature to the observed error. The prior on $G$ consists of a uniform prior between $G_{min}$ and $G_{max}$. Although there is no well defined prescription  for setting a prior on intrinsic scatter, usually a Jeffreys prior is used since it is scale invariant~\cite{Trotta}. Sometimes the choice of prior could also affect the final result.~\cite{Trotta}.  For this work, to be conservative R22 considered multiple choices of priors to understand the impact following some of the considerations in \cite{Gelman06}. 

For the prior on intrinsic scatter, three different priors were used:

\begin{itemize}
\item \textbf{Uniform prior}: This corresponds to a uniform distribution over $\Sigma$ 
\item \textbf{Jeffrey's prior}: This is equivalent to a uniform distribution over $\log \Sigma$
\item \textbf{Inverse Gamma distribution}: The following  function was used for the prior on $\Sigma$.
\begin{equation}
 p(\Sigma|\alpha,\beta) = \frac{\beta^\alpha}{\Gamma(\alpha)}\Sigma^{-(\alpha+1)}\exp[-\frac{\beta}{\Sigma}]
 \label{eq:ig_dist}
\end{equation}
where $\alpha$ and $\beta$ satisfy $\alpha>0$ and $\beta>0$. Both $\alpha$ and $\beta$ are assigned uniform values between 0 and 100.
\end{itemize}

In addition to the aforementioned methods, R22 also used  a  non-parametric method based on Hierarchical Dirichlet Process Gaussian Mixture model (HDPGMM)~\cite{Rinaldi21}  to obtain the central estimate for $W$  mass. The DPGMM is an infinite weighted sum of Gaussian mixture models with Dirichlet  prior being used for the weights~\cite{Fergusen}.
\begin{equation}
    p(x) \approx \sum^{\infty}_i w_i \mathcal{N}(x|\mu_i,\sigma_i)\,.
\end{equation}
In the standard DPGMM, one reconstructs an outer probability distribution from the samples $\mathbf{x} = \{x_1,\ldots,x_N\}$.
However, when we have $N$ sets of inner samples drawn from the posterior distributions,
  one needs to posit a model for both the outer as well as inner posterior samples. This is the 
  central idea behind the HDPGMM. R22 have argued that the likelihood in Eq.~\ref{eq:likelihood} can be interpreted as a DPGMM with a single component $w_i=1$ with every other Gaussian component having $w_j=0$. This model can then be applied to infer the posterior distribution for $G$. More details about  the application of HDPGMM can be found in R22 and Ref.~\cite{Rinaldi21}.

For the parametric Bayesian inference, the posteriors were obtained using {\tt CPnest}~\cite{CPnest} which is based on the nested sampling algorithm. The HPDGMM analysis was done using the {\tt FIGARO} code~\cite{figaro}.
  R22 subsequently applied all the aforementioned four parameter estimation methods to infer a mean value for $G$ along with the intrinsic scatter.

\section{Results}
\label{sec:results}
\subsection{PDG measurements}
The 2022 edition of PDG has collated a total of 25 measurements of the $W$  mass, without the 2022 CDF result. Out of these, only the top 10 measurements were used for the PDG world average of $80.377 \pm 0.012$ GeV.  The complete list of measurements along with the April 2022 CDF result can be found in Table~\ref{tab:PDG}. The last column indicates whether the data was used for calculation of the world average.  More details about each of these measurements can be found in PDG and the references therein.

\subsection{Analysis}
In order to obtain  a central estimate for $M_W$ and $\Sigma$, we need  to select priors for the $M_W$ and $\Sigma$. 
For $M_W$, we use  uniform priors between 80.1 and 80.9 GeV. All the PDG measurements which were used for calculating the world average can be encompassed within this range. For $\Sigma$, the prior depends on the choice of hypothesis followed. We follow the same prescription as in R22.
For the prior on $W$ mass, we have used a uniform prior between the minimum ($m_{min}$) and maximum mass ($m_{max}$) value, depending on the dataset used. For $\Sigma$ we used multiple prior choices, similar to R22. For uniform and Jeffrey's prior the upper bound on $\Sigma$ is given by $0.5 (m_{max}-m_{min})$.  
A summary of all our priors used  can be found in Table~\ref{tab:priors}. We present our results with different combinations of measurements. We describe the annotations we use for these results below:
\begin{enumerate}
\item \textbf{Case I}:  2022 CDF result~\cite{CDF22} (and without the previous 2012 CDF result~\cite{CDF2012})  + all other PDG measurements
\item  \textbf{Case II}: 2022 CDF result~\cite{CDF22} (and without the previous 2012 CDF result~\cite{CDF2012}) + only those PDG measurements which were used for calculation of world average. In other words all the measurements labeled as "not used" were skipped in Table~\ref{tab:PDG}.
\item  \textbf{Case III}: All PDG measurements without the 2022 CDF result. However, instead of the 2022 CDF result~\cite{CDF22}, we use the previous  2012 CDF result~\cite{CDF2012}.
\item  \textbf{Case IV}: Same as Case III, but only those PDG measurements used for world average by PDG.
\end{enumerate}

\begin{table}[!htb]
    \centering
    \begin{tabular}{|P{5cm}|P{5cm}|P{5cm}|}
    \hline
    Model & Parameters & Prior Distribution \\
    \hline
    \SetCell[r=3]{P{5cm}} UNIFORM (UN) &  Mass ($M_W$) & $\mathcal{U}$ [$m_{min}$, $m_{max}$] \\
    & \SetCell[r=2]{P{5cm}} Sigma ($\Sigma$) & $\mathcal{U}$ [$0, \sigma_{max}$] \\
    \hline
    \SetCell[r=3]{P{5cm}} JEFFREY'S (JF) & Mass ($M_W$) & $\mathcal{U}$ [$m_{min}$, $m_{max}$] \\
    & \SetCell[r=2]{P{5cm}} Sigma ($\log \Sigma$) & $\mathcal{U}$ [$0, \sigma_{max}$] \\
    \hline
    \SetCell[r=3]{P{5cm}} INVERSE GAMMA (IG) & Mass ($M_W$) & $\mathcal{U}$  [$m_{min}$, $m_{max}$] \\
    & \SetCell[r=2]{P{5cm}} Sigma ($\Sigma$) & (cf. Eq.\eqref{eq:ig_dist}) \\
    & \SetCell[r=2]{P{5cm}} $\alpha$ & $\mathcal{U}$ [ $0, 100$ ] \\
    & \SetCell[r=2]{P{5cm}} $\beta$ & $\mathcal{U}$ [ $0, 100$ ] \\
    \hline
    \end{tabular} \\
    \caption{Table representing the priors as well as the bounds used for different parameters of the model. In this table, $m_{min}$ and  $m_{max}$ correspond to the minimum and maximum value of  $M_W$ for the dataset considered. Here $\sigma_{max}^2=0.5(m_{max}-m_{min})$. The priors for the parameters for the IG model as well as $\Sigma$ (for all the  models) are same as in R22.}
    \label{tab:priors}
\end{table}
\FloatBarrier

\begin{figure}[!htb]
\centering
\includegraphics[width=10cm, height=10cm]{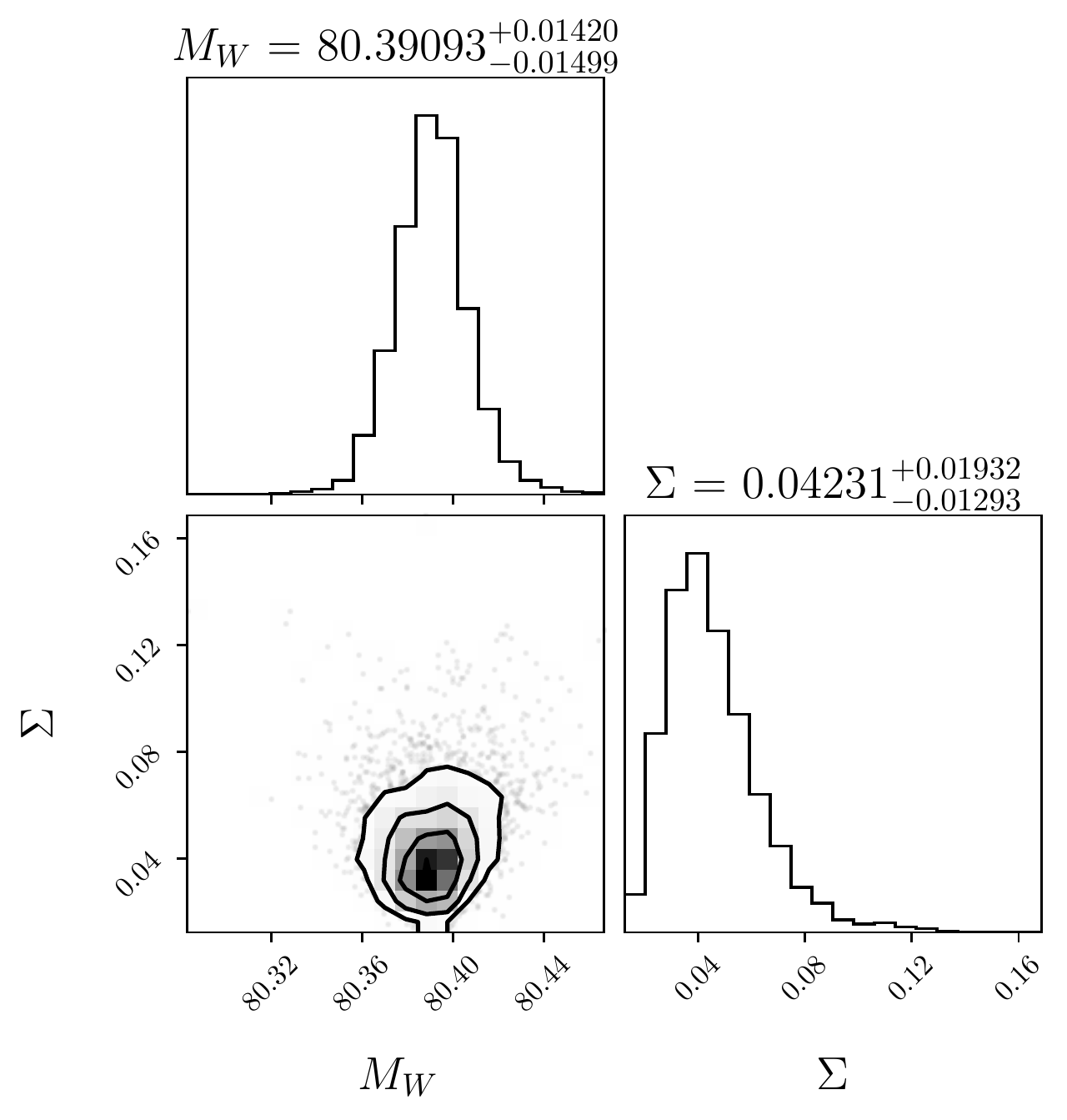}
\caption{\label{fig:UN} Joint posterior distribution for $M_W$ and $\Sigma$ under the UN (Uniform) distribution (Case I). Values above each distribution denote the median value and \rthis{the error bars correspond to the central 68\% credible intervals.} Estimated mass is \textbf{$\mathbf{80.391^{+0.014}_{-0.015}}$ GeV.}}
\end{figure}

\begin{figure}[!htb]
\centering
\includegraphics[width=10cm, height=10cm]{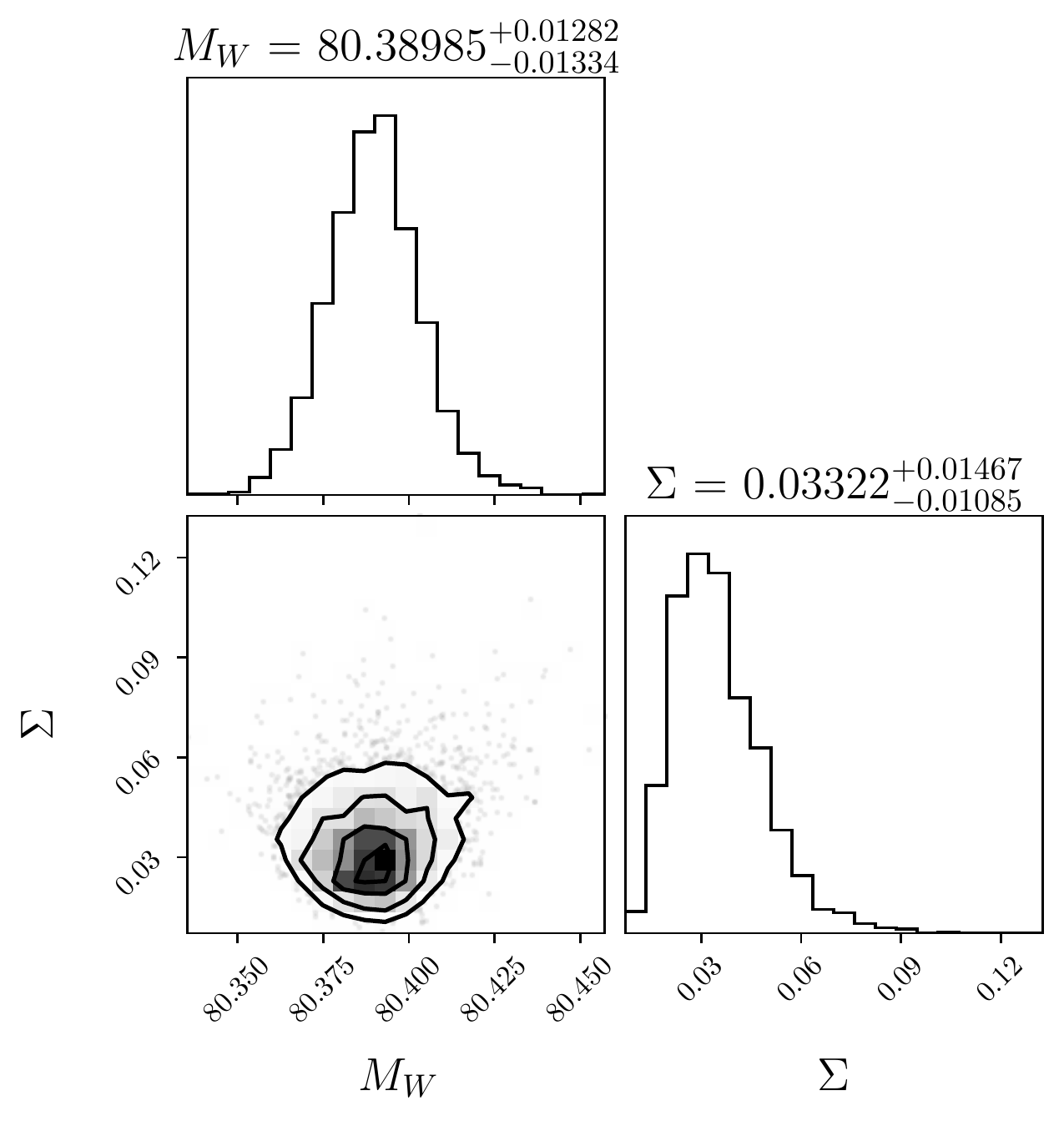}
\caption{\label{fig:JF}Joint Posterior distribution for $M_W$ and $\Sigma$ under the JF (Jeffreys Prior) distribution (Case I). The values above each distribution denote the median value and \rthis{the error bars correspond to the central 68\% credible intervals.} The estimated mass is \textbf{$\mathbf{80.3898 \pm 0.013}$ GeV.}}
\end{figure}

\begin{figure}[!htb]
\centering
\includegraphics[width=10cm, height=10cm]{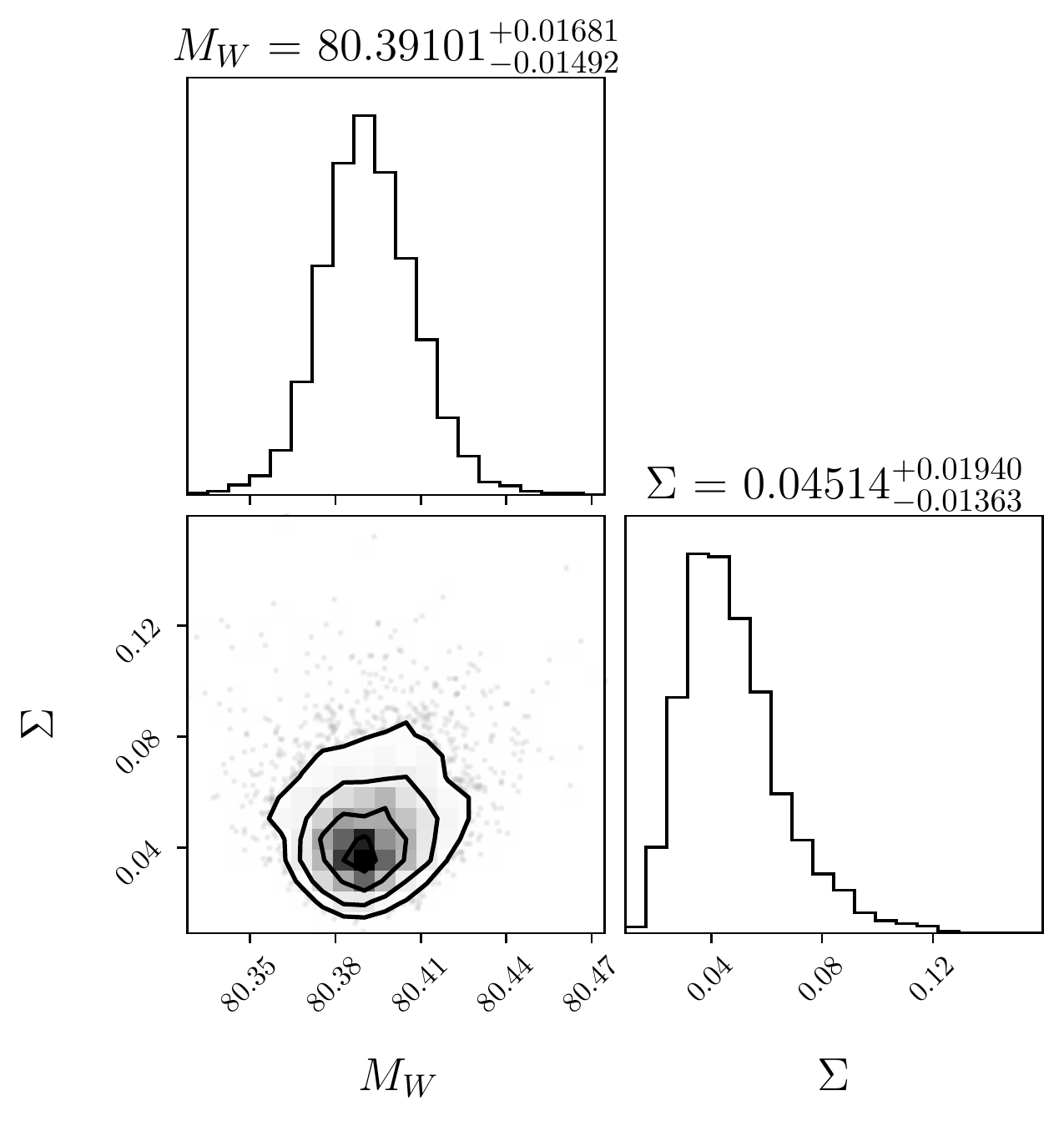}
\caption{\label{fig:IG} Joint Posterior distribution for $M_W$ and $\Sigma$ under the IG (Inverse Gamma) distribution (Case I). The values above each distribution denote the median value \rthis{the error bars correspond to the central 68\% credible intervals.}. The estimated mass is \textbf{$\mathbf{80.391^{+0.017}_{-0.015}}$ GeV.}}
\end{figure}

\begin{figure}[!htb]
    \centering
    \includegraphics[width=10cm, height=10cm]{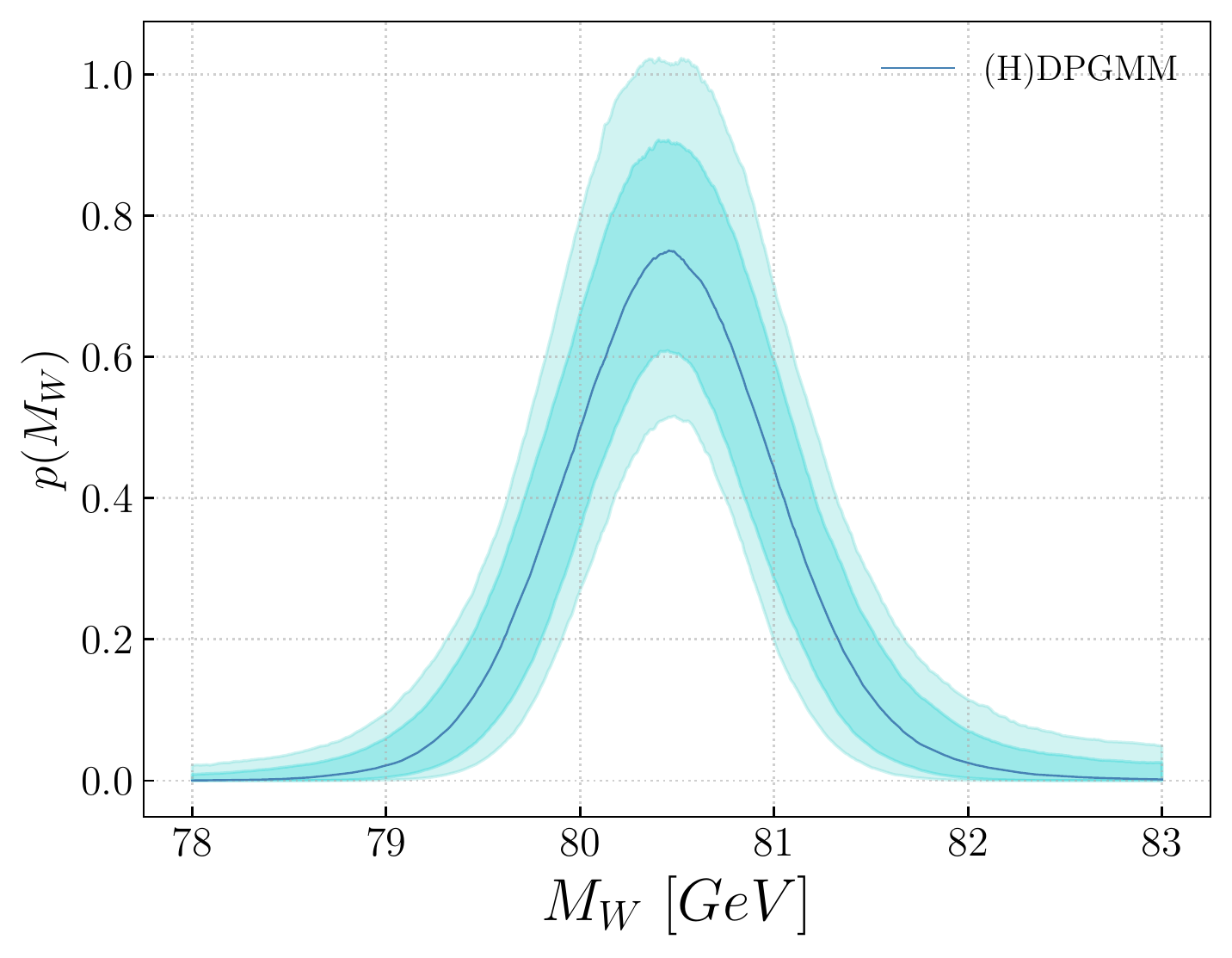}
    \caption{\label{fig:HPDGMM} Posterior distribution for $M_W$ produced by the (H)DPGMM simulation (for Case I). The estimated mass is $\mathbf{80.422^{+0.375}_{-0.322} GeV.}$. The solid blue shows the median mass value, with the 68\% and 90\% credible regions represented by dark and light turquoise, respectively.}
\end{figure}

With the priors in Table~\ref{tab:priors}, we apply the same procedure as in R22 to the aforementioned combinations of datasets. We shall present the  marginalized posterior contours for Case I and present tabular summary of the $M_W$ and $\Sigma$ for all cases.  The marginalized  posterior intervals for $M_W$  and $\Sigma$ for UN, JF, and IG can be found in Figs.~\ref{fig:UN},~\ref{fig:JF}, and ~\ref{fig:IG} respectively. For  HPDGMM, we get a posterior distribution over $M_W$, which can be found in Fig.~\ref{fig:HPDGMM}. The uncertainty  in the $W$-boson estimate has been obtained from the median quantile values from the MCMC chains for each of the posteriors. Therefore, $\sigma_{Mw}$ is not affected by tails in the posterior.
 A tabular summary of  our  results for all the four cases and assumptions on the priors used can be found in Table~\ref{tab:results}. For each value of $M_W$, we also calculate the statistical significance ($N_{\sigma}$), where  $N_{\sigma}$ is given by:
\begin{equation}
        N_{\sigma} = \frac{M_W - 80.357}{\sqrt{(0.006^2 + \sigma_{M_W}^2)}}
        \label{eq:ns}
    \end{equation}
where  $\sigma_{M_W}$  is the uncertainty in $M_W$ and 0.006 GeV is the uncertainty in the Standard Model value~\cite{HillPT}. This is a hybrid estimator of the detection significance combining the tenets of both Bayesian and frequentist statistics. This has been used in Cosmology to quantify the significance of the Hubble tension~\cite{Divalentino} The detection significance $N_{\sigma}$ is a frequentist estimate. However, it uses marginalized central estimates of the W-boson mass which are inferred from  Bayesian analysis. We note that within Bayesian statistics there is no formal way to estimate the $p$-value or a significance.
The value of $N_{\sigma}$ in each of the cases can be found in Table~\ref{tab:my_label}. We can see that if we combine all measurements (for all cases), the discrepancy with respect to the Standard Model value is only 1-2$\sigma$. In Cases I and II where the 2022 CDF result is included, the maximum discrepancy with respect to Standard model is 2.29$\sigma$. The maximum discrepancy is seen for Case III and for JF, where we get a $2.28\sigma$ discrepancy with respect to the Standard model.   For all the cases, HPDGMM shows the least discrepancy, with significance of  less than $1\sigma$ for all the datasets.  The intrinsic scatter varies from 3-6\% depending on the choice of prior and the dataset  considered. This value is much smaller than that estimated by PDG in order to reconcile the 2022 CDF result with previous  measurements. \rthis{We should also point  that the central values in most cases are close to the PDG average. This is most likely due to wide prior range for $\Sigma$.}

\FloatBarrier

\begin{table}[h!]
\centering
\begin{tabular}{|P{2cm}|P{1.75cm}|P{1.75cm}|P{1.75cm}|P{1.75cm}|P{1.75cm}|P{1.75cm}|P{3cm}|}

\hline
& \multicolumn{2}{P{3.5cm}|}{\textbf{UN}} & \multicolumn{2}{P{3.5cm}|}{\textbf{JF}} & \multicolumn{2}{P{3.5cm}|}{\textbf{IG}} & \textbf{(H)DPGMM} \\
\cline{2-8}
& $M_W$ (GeV) & $\Sigma$ & $M_W$ (GeV) & $\Sigma$ &  $M_W$ (GeV) & $\Sigma$ & $M_W$ (GeV) \\
\hline
Case - I & $80.391^{+0.014}_{- 0.015}$ & $0.042^{+0.0193}_{-0.0129}$ & $80.3898^{+0.013}_{-0.013}$ & $0.033^{+0.0147}_{-0.0108}$ & $80.391^{+0.017}_{-0.015}$ & $0.045^{+0.019}_{-0.0136}$ & $80.422^{+0.375}_{-0.322}$ \\
\hline
Case - II & $80.385^{+0.0212}_{-0.021}$ & $0.057^{+0.0291}_{-0.019}$ & $80.386^{+0.0164}_{-0.0174}$ & $0.039^{+0.0205}_{-0.013}$ & $80.385^{+0.0226}_{-0.020}$ & $0.057^{+0.0298}_{-0.0185}$ & $80.384^{+0.050}_{-0.035}$ \\
\hline
Case - III & $80.382^{+0.0122}_{-0.011}$ & $0.0284^{-0.020}_{-0.0145}$ & $80.378^{+0.007}_{-0.007}$ & $0.002^{+0.012}_{-0.0019}$ & $80.382^{0.0138}_{-0.0113}$ & $0.032^{+0.0206}_{-0.0156}$ & $80.377^{+0.035}_{-0.035}$ \\
\hline
Case - IV & $80.377^{+0.0179}_{-0.0160}$ & $0.041^{+0.0276}_{-0.0205}$ & $80.376^{+0.008}_{-0.008}$ & $0.0025^{0.016}_{-0.002}$ & $80.376^{+0.0179}_{-0.0166}$ & $0.042^{+0.0296}_{-0.019}$ & $80.379^{+0.045}_{-0.029}$ \\
\hline
\end{tabular}
\caption{\label{tab:results} Summary of our results for $M_W$ for all the four cases, which encapsulate the different combinations of datasets as well as the four choices for each case. The quoted errors correspond to \rthis{the central 68\% (1$\sigma$) credible intervals.}}
\end{table}

\FloatBarrier
\begin{table}[!htb]
    \centering
    \begin{tabular}{|P{3cm}|P{3cm}|P{3cm}|P{3cm}|P{3cm}|}
    \hline
    & \textbf{UN} & \textbf{JF} & \textbf{IG} & \textbf{(H)DPGMM} \\
    \hline
    Case - I & 2.10$\sigma$ & 2.29$\sigma$ & 1.89$\sigma$ & 0.17$\sigma$ \\
    \hline
    Case - II & 1.27$\sigma$ & 1.57$\sigma$ & 1.20$\sigma$ & 0.54$\sigma$ \\
    \hline
    Case - III & 1.84$\sigma$ & 2.28$\sigma$ & 1.66$\sigma$ & 0.56$\sigma$ \\
    \hline
    Case - IV & 1.06$\sigma$ & 1.90$\sigma$ & 1.01$\sigma$ & 0.48$\sigma$ \\
    \hline
    \end{tabular}
    \caption{\label{tab:my_label} Statistical significance the discrepancy of $M_W$ compared to the Standard  Model value of $M_W=80.357 \pm 0.006$ GeV computed using Eq.~\ref{eq:ns}.}
 \end{table}

 \begin{figure}[!htb]
\centering
\includegraphics[width=10cm, height=10cm]{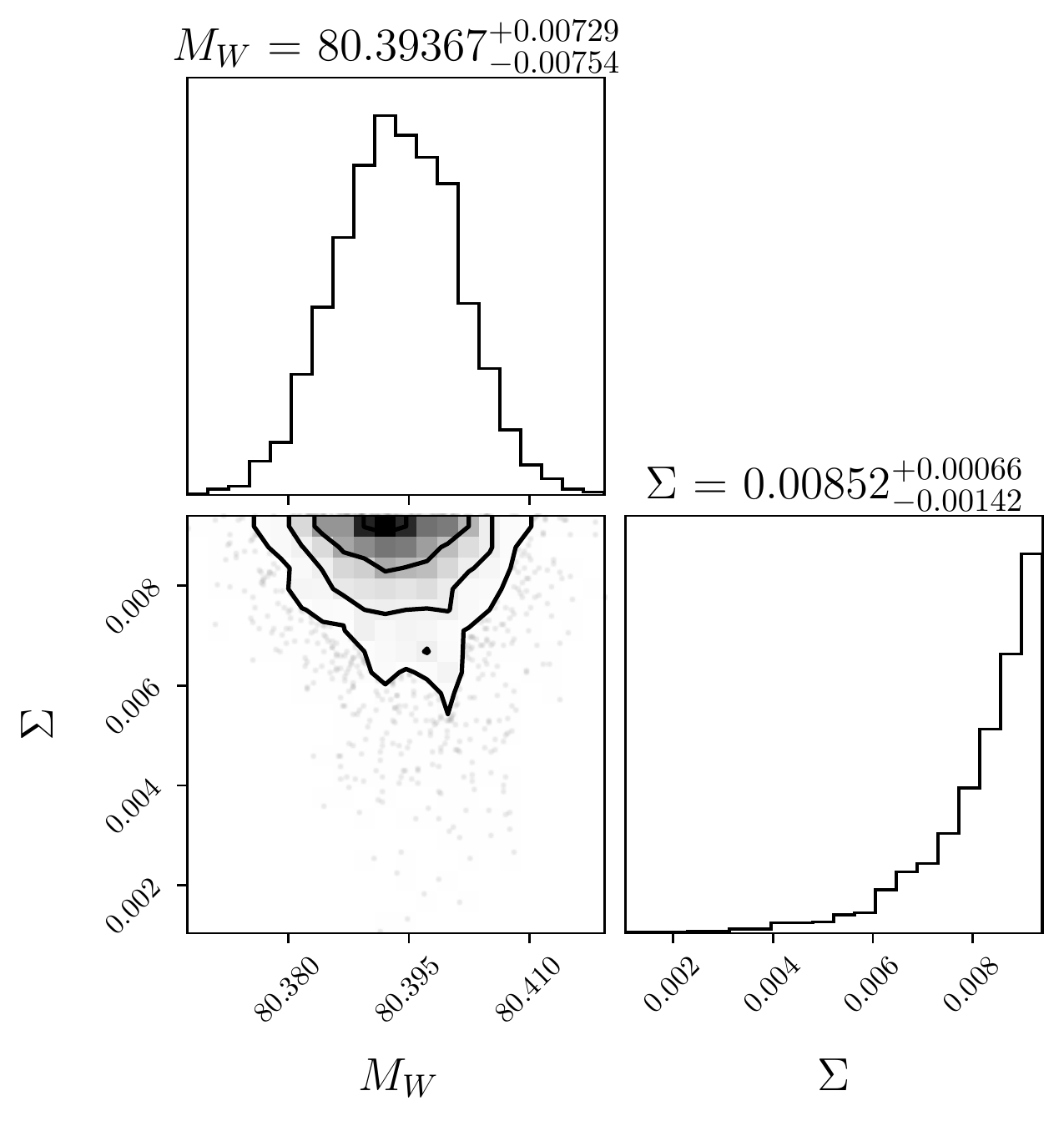}
\caption{\label{fignarrowprior}Joint posterior distribution for $M_W$ and $\Sigma$ under the UN (Uniform) distribution on $[0, 0.0094]$. Values above each distribution
denote the median value and 69\% credible intervals. Estimated mass is \textbf{$\mathbf{80.3937^{+0.0073}_{-0.0075}}$ GeV.}}
\end{figure}

We should point out that one possible reason for the consistency with the Standard Model value could be due to using very broad priors on $\Sigma$. To address this issue, we redid our analysis for the Case-II data (which uses the 2022 CDF result along with other measurements used by PDG for its central estimate), but using a narrow uniform prior on $\Sigma$, given by $\Sigma \in \mathcal{U}$ [0  0.0094]. The upper bound is close the error  reported for the 2022 CDF measurement~\cite{CDF22}.  The posterior distribution for $M_w$ and $\Sigma$ in this case can be found in Fig.~\ref{fignarrowprior}. We noticed that we do not get closed contours for $\Sigma$. So prima-facie, we can only get a lower limit on the intrinsic scatter.  However, since we get a bounded marginalized posterior interval for $M_W$ given by $M_W=80.39367 \pm 0.00754$, we can estimate the detection significance using Eq.~\ref{eq:ns}, which is equal to 3.8$\sigma$. Therefore, we conclude that we if we use a very narrow prior on $\Sigma$, the discrepancy with respect to the Standard Model value remains, but is about 3.8$\sigma$ and is smaller than \rthis{5.77$\sigma$}.

\rthis{Finally, we should point out one caveat with our analysis. For our procedure, we have assigned   the same systematic error $\Sigma$ to all the  measurements. Although Bayesian inference does have the provision for allowing for assigning unique  intrinsic scatter for  individual measurements~\cite{Hogg10}, such an approach would drastically increase the number of free parameters leading to degeneracies between the parameters, when using any MCMC sampler. That is why we have not attempted to do this here. We should however note that our  procedure is similar in spirit although complementary to the PDG analysis, where they have multiplied individual measurement errors by two to reconcile the discrepant sets of measurements~\cite{PDG}. However, one limitation with both these approaches is that one cannot tell which   particular experiment contributes or dominates  towards the extra  scatter that is obtained, as these techniques treat all experimental measurements on an equal footing. However, one would expect that the latter measurements (for eg. the 2022 CDF result) would  have put much more care in controlling systematic uncertainties compared to the earlier ones. Therefore, even though prima-facie  we get values for  $\Sigma$  between 30-60 MeV and the discrepancy with the Standard Model expectation reduces to roughly $0.5-2\sigma$ (depending on the choice of prior), this in no way implies that  the 2022 CDF result has this extra uncertainty of 30-60 MeV   (compared to its quoted value of 6.9 MeV).}

\section{Conclusions}
\label{sec:conclusions}
In April 2022, the CDF collaboration obtained a value of $W$-boson mass given by $M_W=80.4335 \pm 0.0094$ GeV~\cite{CDF22}, resulting in a 7$\sigma$ discrepancy with respect to the Standard Model value and also a 3.7$\sigma$ discrepancy with respect to the 2022  world average  estimated by the PDG (prior to the 2022 CDF result). We  obtain an independent central estimate of $M_W$ from the PDG compilation of the mass measurements, using Bayesian analysis by emulating the same procedure as R22, which was recently  used to obtain a central estimate for $G$ and associated systematic errors. For this purpose, we use a parametric Bayesian model which consists of a Gaussian likelihood and three different  priors on the systematic error as well as a non-parametric method based on HDPGMM.  We also consider different combination of the datasets, with and without the 2022 CDF result.

Our world average  for $M_W$ in all these cases can be found in Table~\ref{tab:results} and the statistical significance with respect to the Standard Model in Table~\ref{tab:my_label}.  We find that for all the choices of priors and combinations of datasets, the discrepancy with respect to the standard model value is less than 3$\sigma$, with the maximum difference being 2.3$\sigma$. We do not find a large difference in the world average between the datasets considered with and without the 2022 CDF result. The intrinsic scatter which we obtained from our analyses varies from 3-6\%. Therefore, there is no statistically significant discrepancy of $W$ boson mass ($>3\sigma$) with respect to the Standard model when we get a world average from all the PDG measurements and use priors on the intrinsic scatter determined by the \rthis{maximum and minimum value of $M_W$ for a given dataset.}

Finally, in order to check if the consistency with the Standard model was because of using a broad prior, we redid our analysis by using a very narrow prior on $\Sigma$ with the upper bound given by the error in the 2022 result, viz 0.0094 GeV. With this prior, we get a discrepancy of 3.8$\sigma$ with respect to  the Standard Model. Therefore, we conclude that with a narrow prior on $\Sigma$,  although the discrepancy  with respect to the Standard model value still persists it is not as pronounced \rthis{as 5.77$\sigma$, obtained   by combining the averaged CDF result with the rest of PDG measurements}. \rthis{However one limitation of our method is that this Bayesian inference technique  does not point out which individual measurement contributes to the extra intrinsic scatter of 30-60 MeV inferred from our analysis. }

We should also point that there are also other Bayesian methods (for eg. methods which allow for outlier rejection~\cite{Hogg10} as well Hierarchical Bayesian analysis~\cite{Sharma} which could be used to get a central estimate of $M_W$ in a complementary way compared to the method used here. We shall pursue this in a future work.

\begin{acknowledgements}
We are grateful to Stefano Rinaldi for making the analysis codes in R22 publicly available, helping us run these codes and also patiently explaining the methodology used in R22 to us. We also thank the anonymous referee for several constructive feedback on the manuscript. 
\end{acknowledgements}

\bibliography{main}
\end{document}